\documentclass{aa}

\usepackage{psfig}

\def\etal{{et\,al. }}
\def\msun{M$_{\odot}$}

\def\degs{\ifmmode ^{\circ}\else$^{\circ}$\fi}
\def\amin{\ifmmode ^{\prime}\else$^{\prime}$\fi}
\def\asec{\ifmmode ^{\prime\prime}\else$^{\prime\prime}$\fi}
\def\fss{\hbox{$.\!\!^{\rm s}$}}        
\def\farcs{\hbox{$.\!\!^{\prime\prime}$}}  
\def\farcm{\hbox{$.\mkern-4mu^\prime$}}    
\def\fm{\hbox{$.\!\!^{\rm m}$}}            
\def\fd{\hbox{$.\!\!^{\rm d}$}}            
\newbox\grsign \setbox\grsign=\hbox{$>$}
\newdimen\grdimen \grdimen=\ht\grsign
\newbox\laxbox \newbox\gaxbox
\setbox\gaxbox=\hbox{\raise.5ex\hbox{$>$}\llap
     {\lower.5ex\hbox{$\sim$}}}\ht1=\grdimen\dp1=0pt
\setbox\laxbox=\hbox{\raise.5ex\hbox{$<$}\llap
     {\lower.5ex\hbox{$\sim$}}}\ht2=\grdimen\dp2=0pt

\def\lax{\mathrel{\copy\laxbox}}
\def\h{$^{\rm h}$}\def\m{$^{\rm m}$}

\begin{document}

   \thesaurus{06         
              (02.01.2;  
               08.02.1:  
               08.02.3;  
               08.09.2:  
               13.25.5)} 

   \title{S 10947 Aql $\equiv$ RX J2009.8+1557: \\ A probable
        RS CVn star which sometimes stops its eclipses\thanks{Table 1
        is available in electronic form
     at the CDS via anonymous ftp to cdsarc.u-strasbg.fr (130.79.128.5)
     or via http:/$\!$/cdsweb.u-strasbg.fr/Abstract.html}}
   \titlerunning{S 10947 Aql: a probable RS CVn star}

   \author{G.A. Richter\inst{1}, J. Greiner\inst{2}}

   \offprints{J. Greiner, jgreiner@aip.de}

   \institute{Sonneberg Observatory, D-96515 Sonneberg, Germany
             \and
             Astrophysical Institute
             Potsdam, An der Sternwarte 16, D-14482 Potsdam, Germany}

   \date{Received 9 May 2000; accepted 17 August 2000}

   \maketitle 

   \begin{abstract}
      We report the discovery of a new variable star, called S 10947 Aql, 
    as the likely optical counterpart of RX J2009.8+1557. The
    optical variability pattern as well as the detected X-ray emission 
    suggest that it is a chromospherically active binary of the 
    RS Canum Venaticorum ($\equiv$ RS CVn) type. 
    We discovered an occasional disappearance
   of the eclipsing minima as well as large variations in the eclipse
   amplitude. We discuss possible causes of this peculiarity.


   \end{abstract}

\section{Introduction}

RS Canum Venaticorum (RS CVn) variables are described by Hall (1972;
see also Biermann \& Hall 1976), as close binaries comprising a 
G- or K-type subgiant and a F- or G-type star of luminosity class IV--V
with the following special property (see also Zeilik \etal\ 1979): 
Their light curves are characterized by long waves with amplitudes 
up to 0.2 mag which, if the binary is eclipsing, as a rule move
towards smaller phase of the orbital light curve. This is typically interpreted
as the effect of large star spots on one hemisphere of the cool component
when rotating with a speed slightly different from the synchronous rotation
(e.g. Catalano 1983). Thus, these waves are assumed to be the beat between the
orbital period and the slightly out-of-phase differential rotation 
of the spotted star. The assumption of substantial chromospheric activity 
is supported by additional observational features, such as strong \ion{Ca}{II}
emission lines, lively flaring activity in the optical and other spectral 
regions and variable X-ray emission (e.g. Walter \etal\ 1980, Charles 1983).

\section{Observations and Discussion}

As part of our programme of investigating the optical long-term behaviour of
selected ROSAT sources (e.g. Richter \etal\ 1995)
we discovered a new variable star as the likely optical counterpart
of RX J2009.8+1557. This object which we called S 10947 Aql, 
varies in the B band between 13\fm4 and 14\fm8. It is identical to 
GSC 0161801655.
The optical coordinates are:

\centerline{R.A. = 20\h09\m51\fss43, Dec. = +15\degs57\amin34\farcs2}

\noindent (equinox 2000.0), consistent with our measurement on the
digitized Palomar Sky Survey plate (see marked object in Fig. 1).

\begin{figure}[bh]
 \vbox{\psfig{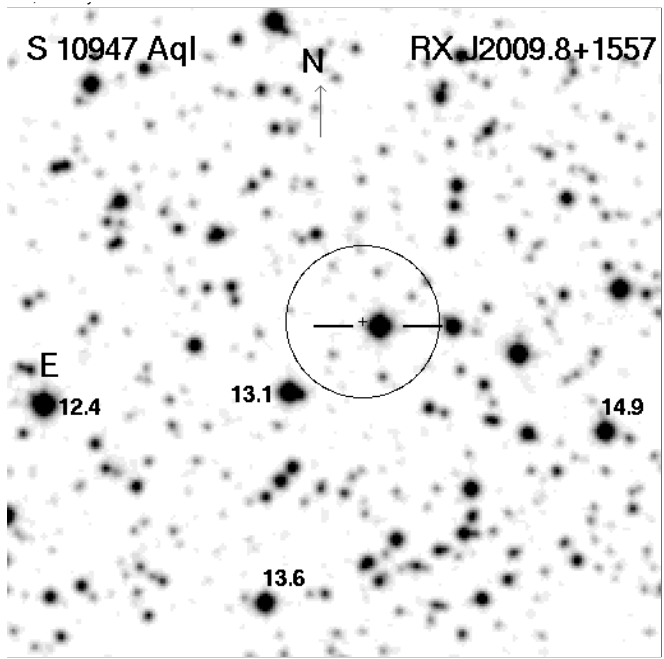}}\par
 \caption[fc]{A 4\farcm3 by 4\farcm3 part of the
  digitized sky-survey image (based on the red passband plate SF04843 taken on 
 25 Sep 1992) with the X-ray error circle of the ROSAT all-sky survey position 
  (large circle; 30\asec\ radius) overplotted. S\,10947 Aql is marked by two 
  heavy dashes. The numbers next to 4 other bright stars are photographic
  magnitudes of the comparison stars (see Tab. \ref{calib})
  used for deriving the light curve of 
  S\,10947 Aql (see Fig. 2). They have been derived by differential
  connection to comparison stars of WZ Sge (Khruzina \& Shugarov 1991). }
\end{figure}

\begin{figure}
  \vbox{\psfig{figure=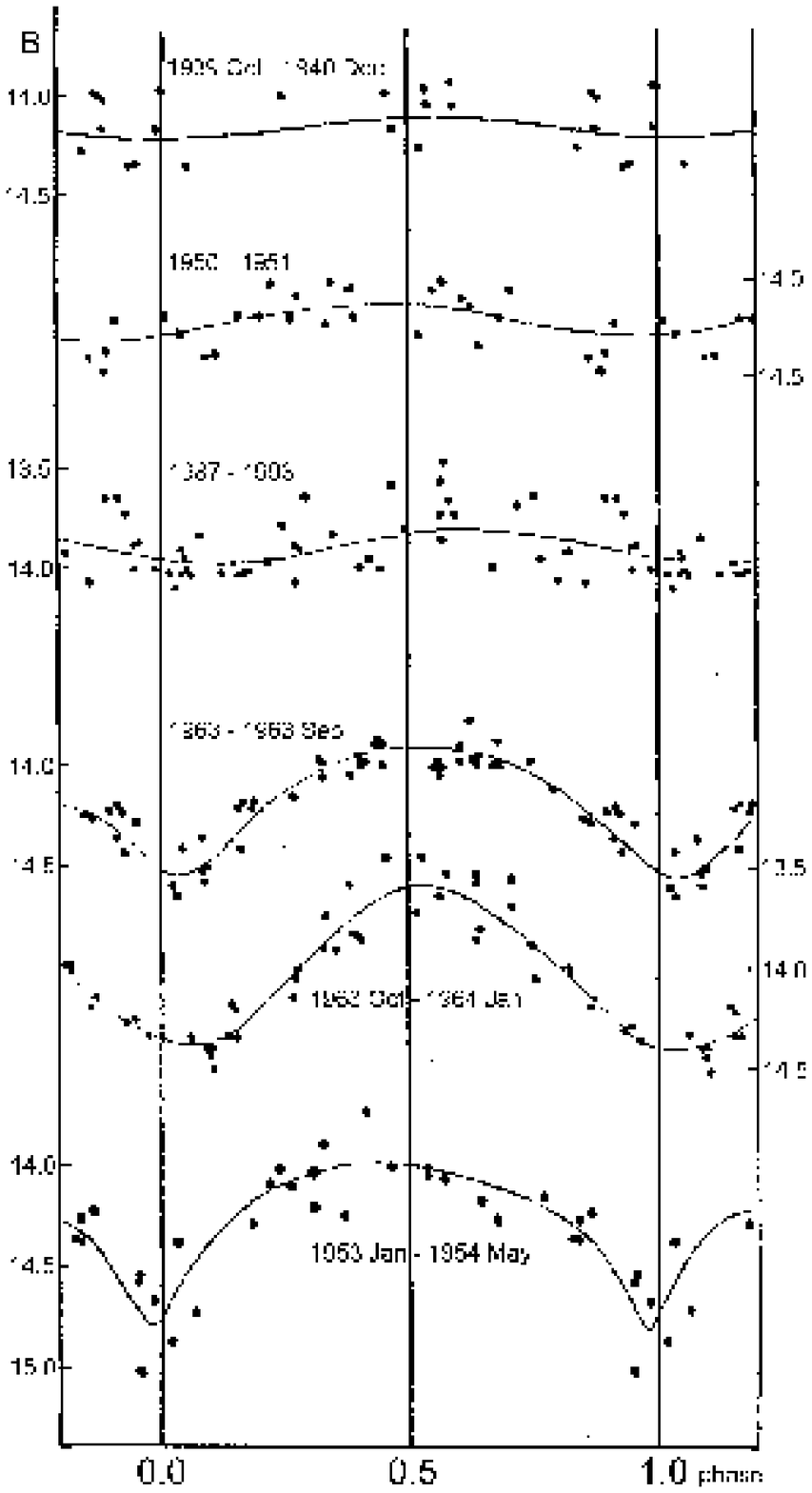,width=8.7cm}}\par
\caption[ltlc]{Some examples of light curves without eclipses (upper half)
   and with eclipses (lower half). Data taken over several month (time span 
   is indicated above each light curve) are folded over P$_{\rm h}$.}
\end{figure}

RX J2009.8+1557 was detected during the ROSAT all-sky survey in 1990
at a count rate
of 0.019$\pm$0.005 cts/s with a likelihood of detection of 12.4 (corresponding
to about 4$\sigma$ confidence). With a total of only 9 counts collected 
during the total exposure time of 495 sec this source is below the
brightness limit of the 1 RXS catalog (Voges \etal\ 1999). 
Nonetheless, these few  photons supply both  a well-defined position of
RA (2000.0) = 20\h 09\m 51\fss9, Decl. (2000.0) = +15\degs 57\amin 35\asec\
with an error radius of 30\asec\ 
as well as an indication for an absorbed, hard X-ray spectrum (in ROSAT 
standards). The hardness ratios are 
HR1 = 
(N$_{\rm 52-201}$ -- N$_{\rm 11-41}$)/(N$_{\rm 11-41}$ + N$_{\rm 52-201}$) 
=  0.70$\pm$0.32, and HR2 = 
(N$_{\rm 91-200}$ -- N$_{\rm 50-90}$)/N$_{\rm 50-200}$ = 
0.52$\pm$0.34,
where N$_{\rm a-b}$ denotes the number of counts in ROSAT's position 
sensitive proportional counter between
channel a  and channel b).
Adopting a Raymond-Smith spectrum with a 1 keV temperature and an absorbing
column of $N_{\rm H}$=1$\times$10$^{21}$ cm$^{-2}$ (corresponding to
$\sim$50\% of the total column of $N_{\rm H}$=1.95$\times$10$^{21}$ cm$^{-2}$ 
in this direction; Dickey \& Lockman 1990)
we derive an unabsorbed X-ray intensity of 
$L(0.1-2.4\ {\rm keV}) = 1.8 \times 10^{-13}$ erg cm$^{-2}$ s$^{-1}$.
This corresponds to an emission measure of 
1.3$\times$10$^{52}$ (D/100 pc)$^2$ cm$^{-5}$. 
No ROSAT pointing exists for this sky area which could allow to derive better
constraints.

\begin{figure}
   \vbox{\psfig{figure=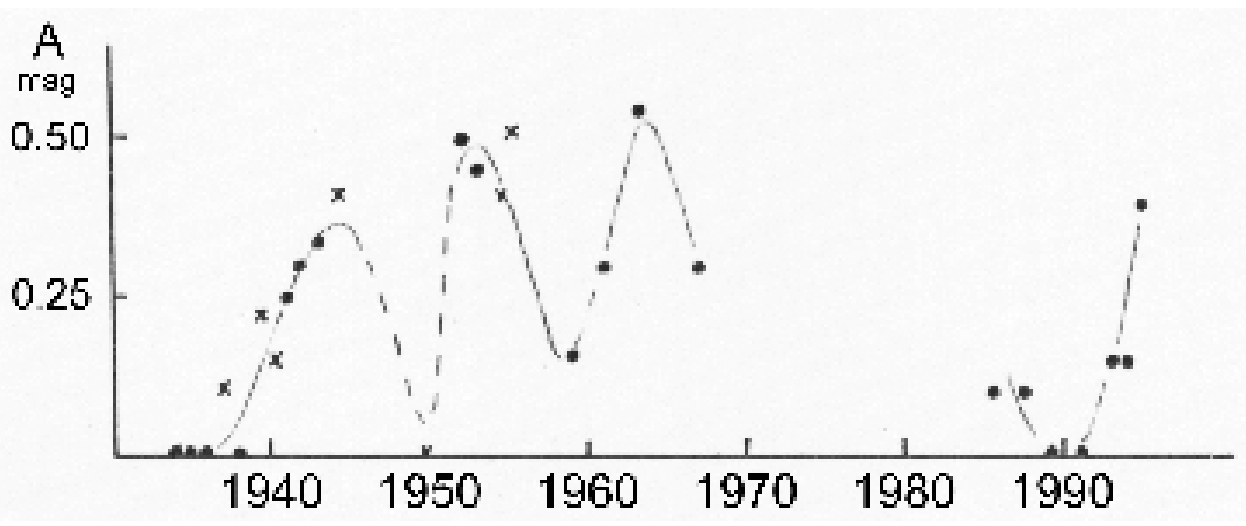,width=8.7cm}}\par
\caption[ecl]{Time variation of the eclipsing amplitude. The time axis is 
    in years, and crosses denote uncertain measurements.}
\end{figure}

Altogether 709 archival plates taken between 1934 and 1995 of the Sonneberg
astrographs 400 mm, 170 mm and 140 mm were used for investigating the 
long-term behaviour of S 10947 Aql. Tab. \ref{phot} gives the details of these
measurements. Unfortunately, the object is invisible
on all plates of the Sonneberg sky patrol.

Spectroscopic observations of S 10947 Aql do not exist, but the kind of 
brightness changes together with the detected X-ray emission seem to indicate 
that it is a chromospherically active binary of the 
 RS CVn type.

\begin{figure}
  \vbox{\psfig{figure=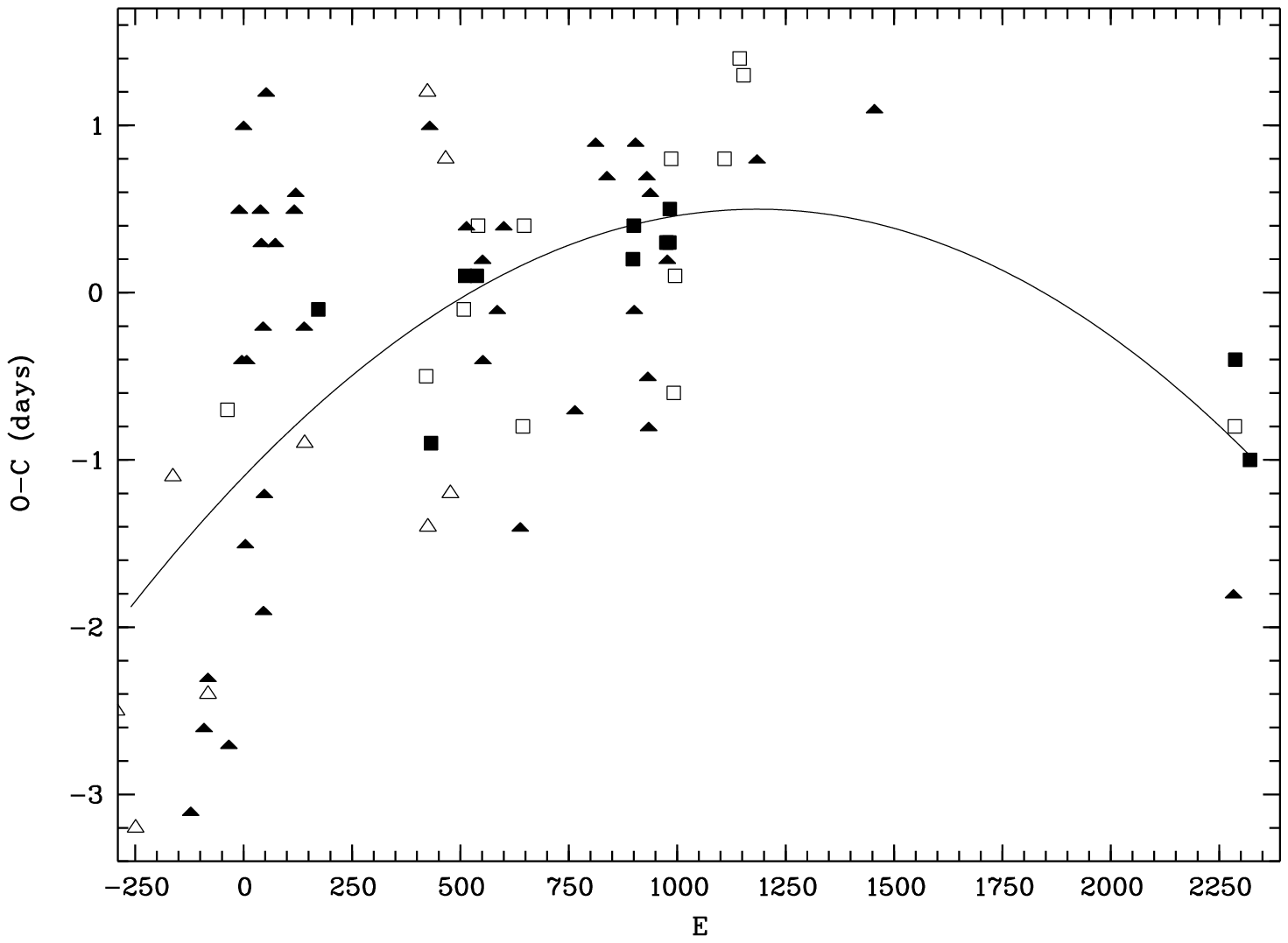,width=8.7cm}}\par
  \vspace{-0.25cm}
\caption[o-c]{$O-C$ diagram, calculated according to the formula
   242\,9812.40 + 8\fd6294$\times E$. The solid line is the best-fit
   quadratic adaption to the observations: 
 $m$ = 242\,9811.30 + 8\fd6321 $\times E$ - 1.14$\times$10$^{-6}$ $\times E^2$.
  The symbols have the following meaning:
  filled triangles = secure isolated observations (error $\pm$0\fd7), 
  open triangles = uncertain isolated observations (error $\pm$1\fd5),
  filled squares = secure mean of series observations (error $\pm$0\fd2),
  open squares = uncertain mean of series observations (error $\pm$0\fd6).
  }
  \label{o-c}
\end{figure}

The major results of the variability study can be summarized as follows:
\begin{itemize}
\vspace{-0.25cm}\item We find occasional brightness humps which can be 
 recognized and 
 followed over  several months to years, with a cycle length of about 
 P$_{\rm h}$ = 8\fd628 (in the mean) and an amplitude of $\lax$ 0.5 mag
 (migrating waves).
\item Eclipsing minima occur with a mean period of 
  P$_{\rm o}$ = 8\fd6294.
\item We observe the occasional disappearance
  of the eclipsing minima (see Fig. 2 and 3). This is probably only partly
  caused by the overlapping of the minima with the maxima of the migrating
  waves. Also, in some other RS CVn stars the depth of the primary minima 
  varies, but it does not approach zero as is the case in   S 10947 
  (e.g. RT Lac: the amplitude varies between about 0.7 and 1.0 mag,
  see Evren 1989; 
  UX UMa: the amplitude varies from 0.4 to about 0.6 mag, see Geyer 1980).
\item 
 During the whole time of observation the period P$_{\rm o}$ is decreasing
 (Tab. \ref{omc} and Fig. \ref{o-c}). The observed minima may formally be
 described by
 $$ m = 242\,9811.30 + 8\fd6321 \times E - 1.14 \times 10^{-6} \times E^2 $$
 The period of most, if not all, RS CVn stars is changing, but in our case 
 the numerical value of the quadratic term is quite large (see below). 
Of course, it is not known whether the 
 periods are in reality changing smoothly, or whether the changes are
  abrupt (polygonal).
\end{itemize}

\begin{table}
  \caption{Photometric observations of S 10947 Aql at Sonneberg Observatory.
     Only the first (seven) and (five) latest measurements are given to 
    indicate the temporal coverage. The number after each plate type gives 
    the total number of plates investigated. The full table is available 
    electronically at CDS.}
  \scriptsize
  \vspace{-0.1cm}
  \begin{center}
  \begin{tabular}{ccc}
  \hline \noalign{\smallskip}
    HJD (2400000+) &     m$_{\rm pg}$ (mag)  & Uncert. flag$^{(1)}$ \\
   \noalign{\smallskip} \hline \noalign{\smallskip}
   \multicolumn{3}{c}{\bf A plates (175)} \\
 27543.564 & 13.55 & \\
 27546.563 & 13.83 & \\
 27569.510 & 13.63 & \\
 27579.540 & 14.05 & \\
 27612.513 & 13.67 & \\
 27628.505 & 13.59 & \\
 27635.443 & 13.95 & \\
   ...     & ...   & \\
 39765.310 & 13.69 & \\
 40030.462 & 14.35 & \\
 40059.411 & 13.60  &   : \\
 40415.454 & 13.59 & \\
 40477.337 & 13.82  & \\
\noalign{\smallskip}
\multicolumn{3}{c}{\bf D plates (140)} \\
 33772.546 & 14.16 & \\
 33778.542 & 14.03 & \\
 33809.521 & 14.10  &    : \\
 33828.450 & 14.12 & \\
 33834.495 & 14.36 & \\
 33838.426 & 14.00  &  : \\
 33855.432 & 14.22 & \\
   ...     & ...   & \\
 35571.609 & 13.71 & \\
 35609.550 & 13.82 & \\
 36343.450 & 14.30  &   : \\
 36347.523 & 14.30  &  :: \\
 36404.542 & 14.80  &  :: \\
\noalign{\smallskip}
\multicolumn{3}{c}{\bf F plates (96)} \\
 26928.422 & 14.10  &    :: \\
 26931.487 & 14.30  &    : \\
 26980.338 & 13.90  &    : \\
 27003.276 & 13.93 & \\
 27277.441 & 14.00  &       : \\
 27281.425 & 14.30  &       : \\
 27298.394 & 13.90  &       :: \\
   ...     & ...   & \\
 37246.220 & 13.60 &    : \\
 37559.390 & 14.00 & \\
 37824.515 & 13.38 & \\
 37934.428 & 14.00  &  : \\
 39352.395 & 13.89  & \\
\noalign{\smallskip}
\multicolumn{3}{c}{\bf GA, GB, GC plates (298)} \\
 29054.501 & 14.01 & \\
 29102.537 & 14.35 & \\
 29130.461 & 14.06 & \\
 29162.322 & 14.17 & \\
 29168.367 & 13.92 & \\
 29429.518 & 14.18 & \\
 29438.498 & 14.11 & \\
   ...     & ...   & \\
 49862.479 & 13.59 & \\
 49866.533 & 13.90 & \\
 50246.521 & 13.84 & \\
 50248.516 & 13.68 & \\
 50281.543 & 13.79 & \\
\noalign{\smallskip}
   \noalign{\smallskip} \hline
   \end{tabular}
   \end{center}

   \vspace{-0.25cm}\noindent $^{(1)}$ The symbols have the following meanings: 
         ``:''  $\equiv$ uncertain,
         ``::'' $\equiv$ very uncertain,
         ``$>$''  $\equiv$ upper limit.
   \label{phot}
   \end{table}

According to the classical interpretation of RS CVn stars the brightness
changes can be interpreted solely by starspot activities in a binary system
(e.g. Geyer 1976). But until now it was not yet possible to unambiguously 
explain the physical processes in RS CVn systems. This is because
photoelectric and spectroscopic observations are not available to a sufficient
extent since the phenomenon was discovered by Hall (1972).

Systematic changes of the orbital period are
observed also in most other RS CVn systems.
Hall \& Kreiner (1980) and Hall \etal\ (1980) 
gave a compilation of 34 such objects where both, 
decreasing and increasing periods are found in the ratio of about 2:1.
The value of d\,log $P$/d$t$ = --2.3 $\times$ 10$^{-6}$ for S 10947
is large but not extraordinary. It is surpassed only by 
SZ Psc (--5.25 $\times$ 10$^{-6}$), CQ Aur (--2.45 $\times$ 10$^{-6}$)
and AR Mon (--1.22 $\times$ 10$^{-6}$).

In any case, large period changes are an indication of rapid evolutionary
efects (e.g. p. 427 in Kopal 1978). However, it is difficult to estimate
more details, such as mass loss or mass transfer rates in RS CVn stars
because the period changes may be caused by effects which are not directly
related to the mass transfer in the binary system. As far as known, RS CVn
stars have binary components of similar mass. Though RS CVn stars
are detached systems, mass exchange, if occuring, does not much influence
the period length. Though Evren (1989) shows that in the case of RT Lac
an intermittent gas flow seems to exist between the binary components,
additional quantities must be known to answer this question with more
confidence. Hall (1972), for example, proposes that ejection of matter 
in regions of very strong starspot activity on one side of the cooler component
may give large contributions to the changes in orbital period (rocket effect).
Sahade \& Wood (1978, on page 66) write that ``the large period changes
found in the RS CVn variables are still not well understood ... and the energy
required to eject larger amounts of mass must be very large''.
Other authors (e.g. Kopal 1978) state that it is difficult to estimate 
the mass loss. Nevertheless, Hall \& Kreiner (1980) and Hall \etal\ (1980) 
tried to give crude 
estimates of the range of mass loss rates in 34 RS CVn-type stars
based on the hypothetical assumptions of magnetic field strengths
and further quantities. They got values up to about 10$^{-6}$ \msun/yr
in some extreme cases.

\begin{table}
\caption{Observed minima and $O-C$ values}
\vspace*{-0.2cm}
\scriptsize
\begin{center}
\begin{tabular}{cccrcc}
\hline \noalign{\smallskip}
I(solated) & HJD   & $m_{\rm pg}$ & E & $O-C$ & Uncert. \\
S(eries)   & 24... &  (mag)       &   &       & flag \\
\noalign{\smallskip} \hline \noalign{\smallskip}
I & 27281.4 & 14.35 & -293 & -2.5 & : \\ 
I & 27660.5 & 14.26 & -249 & -3.2 & : \\ 
I & 28404.4 & 14.30 & -163 & -1.1 & : \\ 
I & 28756.5 & 14.70 & -122 & -3.1 &  \\ 
I & 29024.6 & 14.41 &  -91 & -2.6 &  \\ 
I & 29102.4 & 14.28 &  -82 & -2.4 & : \\ 
I & 29102.5 & 14.35 &  -82 & -2.3 &  \\ 
S & 29492.4 & 14.32 &  -37 & -0.7 & : \\ 
I & 29516.4 & 14.31 &  -34 & -2.7 &  \\ 
I & 29726.6 & 14.34 &  -10 & +0.5 &  \\ 
I & 29777.5 & 14.47 &   -4 & -0.4 &  \\ 
I & 29813.4 & 14.40 &    0 & +1.0 &  \\ 
I & 29846.4 & 14.37 &    4 & -1.5 &  \\ 
I & 29872.4 & 14.34 &    7 & -0.4 &  \\ 
I & 30149.5 & 14.41 &   39 & +0.5 &  \\ 
I & 30166.5 & 14.50 &   41 & +0.3 &  \\ 
I & 30200.5 & 14.41 &   45 & -0.2 &  \\ 
I & 30207.4 & 14.50 &   46 & -1.9 &  \\ 
I & 30225.4 & 14.47 &   48 & -1.2 &  \\ 
I & 30262.3 & 14.71 &   52 & +1.2 &  \\ 
I & 30442.6 & 14.38 &   73 & +0.3 &  \\ 
I & 30520.5 & 14.34 &  117 & +0.5 &  \\ 
I & 30848.5 & 14.50 &  120 & +0.6 &  \\ 
I & 31020.3 & 14.44 &  140 & -0.2 &  \\ 
I & 31028.3 & 14.31 &  141 & -0.9 & : \\ 
S & 31296.6 & 14.65 &  172 & -0.1 &  \\ 
S & 33444.9 & 14.53 &  421 & -0.5 & : \\ 
I & 33472.5 & 14.50 &  424 & +1.2 & : \\ 
I & 33478.5 & 14.40 &  425 & -1.4 & : \\ 
I & 33515.4 & 14.37 &  429 & +1.0 &  \\ 
S & 33539.0 & 14.45 &  432 & -0.9 &  \\ 
I & 33834.5 & 14.36 &  466 & +0.8 & : \\ 
I & 33927.4 & 14.36 &  477 & -1.2 & : \\ 
S & 34196.0 & 14.50 &  508 & -0.1 & : \\ 
S & 34222.1 & 14.50 &  511 & +0.1 &  \\ 
I & 34248.4 & 14.80 &  514 & +0.4 &  \\ 
I & 34334.3 & 14.60 &  524 & +0.1 &  \\ 
S & 34455.1 & 14.70 &  538 & +0.1 &  \\ 
S & 34481.3 & 14.60 &  541 & +0.4 & : \\ 
I & 34567.4 & 14.90 &  551 & +0.2 &  \\ 
I & 34575.5 & 15.20 &  552 & -0.4 &  \\ 
I & 34860.6 & 14.60 &  585 & -0.1 &  \\ 
I & 34990.4 & 14.78 &  600 & +0.4 &  \\ 
I & 35316.5 & 14.60 &  638 & -1.4 &  \\ 
S & 35368.9 & 14.60 &  644 & -0.8 & : \\ 
S & 35396.0 & 14.60 &  647 & +0.4 & : \\ 
I & 36404.5 & 14.80 &  764 & -0.7 &  \\ 
I & 36820.4 & 14.45 &  812 & +0.9 &  \\ 
I & 37044.6 & 14.49 &  838 & +0.7 &  \\ 
S & 37561.8 & 14.60 &  898 & +0.2 &  \\ 
S & 37579.3 & 14.60 &  900 & +0.4 &  \\ 
I & 37587.3 & 14.43 &  901 & -0.1 &  \\ 
I & 37614.3 & 14.78 &  904 & +0.9 &  \\ 
I & 37838.5 & 14.45 &  930 & +0.7 &  \\ 
I & 37854.5 & 14.41 &  932 & -0.5 &  \\ 
I & 37871.5 & 14.35 &  934 & -0.8 &  \\ 
I & 37907.4 & 14.36 &  938 & +0.6 &  \\ 
S & 38226.4 & 14.59 &  975 & +0.3 &  \\ 
S & 38235.0 & 14.71 &  976 & +0.3 &  \\ 
I & 38243.5 & 14.58 &  977 & +0.2 &  \\ 
S & 38286.8 & 14.53 &  982 & +0.3 &  \\ 
S & 38295.6 & 14.47 &  983 & +0.5 &  \\ 
S & 38321.8 & 14.50 &  986 & +0.8 & : \\ 
S & 38372.2 & 14.40 &  992 & -0.6 & : \\ 
S & 38398.8 & 14.40 &  995 & +0.1 & : \\ 
S & 39383.2 & 14.40 & 1109 & +0.8 & : \\ 
S & 39685.8 & 14.20 & 1144 & +1.4 & : \\ 
S & 39763.4 & 14.40 & 1153 & +1.3 & : \\ 
I & 40030.5 & 14.35 & 1184 & +0.8 &  \\ 
I & 42369.2 & 14.40 & 1455 & +1.1 &  \\ 
I & 49511.6 & 14.07 & 2283 & -1.8 &  \\ 
S & 49538.4 & 13.88 & 2286 & -0.8 & : \\ 
S & 49547.4 & 13.93 & 2287 & -0.4 &  \\ 
S & 49840.2 & 14.04 & 2321 & -1.0 &  \\ 
\hline 
\end{tabular}
\end{center}
\label{omc}
\end{table}

As already mentioned, the eclipsing minima in S 10947 occasionally disappear 
(see Fig. 2 and 3). Apart from S 10947 only very few binaries (none of 
which is a RS CVn star) are known to have occasionally vanishing eclipse 
amplitudes, and thus not much is known about their cause.

There are several possible explanations for varying eclipse amplitudes:
\begin{itemize}
\item {\it Triple system}:
An obvious explanation of varying orbital period and eclipse depth is
due to the influence of a third body in orbit (e.g. Africano \& Wilson 1976).
This possibility was later abandoned as a general explanation since 
(i) $O-C$ variations were
actually observed in too many cataclysmic variables to attribute all to third
bodies, and (ii) for the case of UX UMa the apparent sinusoidal $O-C$ 
variation was found not to repeat (Rubinstein \etal\ 1991).
One  exception is SS Lac which has ceased eclipsing completely. This is 
believed to be due to a change
of the orbital inclination angle caused by a third stellar component
which gives not only rise to an apsidal motion of probably $\sim$ 1000 years,
but also to a periodic oscillation of the orbital inclination with a very
long time scale (Milone \etal\ 2000, Torres \& Stefanik 2000).
\item {\it Dust obscuration}:
Irregularities of production and destruction of carbon dust can hide
the minima. As far as we know, this phenomenon primarily leads to
small amplitude variations rather than a change of 0.5 mag as observed in
S 10947.
\item {\it Strong wind}:
Variations of the accretion rate may easily affect variations in the amount
of matter expelled by the system in a wind. As an example, the eclipsing 
binary CV Ser (not a RS CVn star) showed no eclipse light variations at all
in 1970. According to Cowley \etal\ (1977) (see also Hoffmeister \etal\ 1984),
it is not one of the stellar binary components which is eclipsed,
but rather some bright material between the stars.
\end{itemize}

For S 10947, the short time scale and the non-periodic behaviour of the times
of vanishing amplitude seem to rule out the influence of a third component.
We therefore suppose that these eclipses are caused by circumstellar matter
or changes in the accretion disk, variable in size and/or brightness.
This supports earlier statements by Hall \& Ramsey (1992) that extended
matter in RS CVn systems may play a greater role than hitherto assumed.

\section{Conclusions}

Because of 
(1) the small difference to the best-fit X-ray position of 
only $\sim$5\asec\ (far below the X-ray position error),
(2) the lack of any other optical objects within the X-ray error circle 
brighter than 19 mag. and
(3) a X-ray-to-optical luminosity ratio of $L_{\rm X}/L_{\rm opt} \approx 0.01$
which is in the range exhibited by RS CVn stars,
we are quite certain about the identification
of RX J2009.8+1557 with S 10947 Aql.

S 10947 Aql shows
interesting properties in its eclipsing light curve which
may contribute to a better understanding of the RS CVn systems.

This work demonstrates the importance of plate archives where many still 
unknown secrets and information are hidden. Detailed spectroscopic
observations are urgently needed to further the understanding of this enigmatic
source S 10947 Aql. Also, continued long-term monitoring should determine the
further evolution of the $O-C$ curve, and thus can prove/disprove
the influence of a third body.

\begin{acknowledgements}
We are pleased to acknowledge post-facto photoelectric calibration of the
field by Arne Henden (see Tab. \ref{calib}) as suggested by the referee.
We are indebted to P. Kroll (Sonneberg Observatory) for transforming
the photographic measurements into electronic format.
GAR and JG were partly supported by the German Bundes\-mini\-ste\-rium f\"ur 
Bildung, Wissenschaft, Forschung und Technologie (BMBF/DLR) under contract 
05\,2S\,0524 and 50\,OR\,96\,02\,3, respectively.
The ROSAT project is supported by BMBF/DLR and the Max-Planck-Society.
The finding chart (Fig. 1) is
based on photographic data of the National Geographic Society -- Palomar
Observatory Sky Survey (NGS-POSS) obtained using the Oschin Telescope on
Palomar Mountain.  The NGS-POSS was funded by a grant from the National
Geographic Society to the California Institute of Technology.  
The Digitized Sky Survey was produced at the Space
Telescope Science Institute under US Government grant NAG W-2166.
This research has made use of the USNO-A1.0 catalog produced by the
U.S. Naval Observatory.
\end{acknowledgements}

\bigskip\bigskip
\noindent{\it Note added in proof:}

Arne Henden, following our request in response to a suggestion of the referee,
has performed a CCD-based photometric calibration as given below. 

\begin{table}[hb]
  \vspace{-0.25cm}
  \caption{Comparison stars}
  \vspace{-0.2cm}
  \begin{tabular}{ccc}
  \hline \noalign{\smallskip}
    Star position (2000)$^{(1)}$ & $m_{\rm pg}$ (mag) & $B$ (mag)$^{(1)}$ \\
  \noalign{\smallskip} \hline \noalign{\smallskip}
  20\h 09\m 53\fss92~ +15\degs 57\amin 07\farcs8 & 13.10$\pm$0.07 & 13.280$\pm$0.006 \\
  20\h 10\m 00\fss63~ +15\degs 57\amin 02\farcs2 & 12.39$\pm$0.07 & 12.770$\pm$0.005 \\
  20\h 09\m 45\fss25~ +15\degs 56\amin 52\farcs8 & 14.99$\pm$0.07 & 15.040$\pm$0.001 \\
  20\h 09\m 54\fss54~ +15\degs 55\amin 44\farcs0 & 13.63$\pm$0.07 & 13.734$\pm$0.013 \\
   \noalign{\smallskip} \hline
   \end{tabular}

   \noindent $^{(1)}$ The field photometry is based on two photometric 
   nights in June 2000 with the Naval Observatory Flagstaff Station
    1.0 m telescope. A Tektronix 1024*1024 CCD was used with Johnson-Cousins 
   BVRI filters. Typical nightly zero point errors are less than 0.02 mag.
   Astrometry was performed with respect to USNO-A2.0 and has internal 
   errors of less than 100 mas.
    The full calibration file including
    VRI measurements is electronically available from A. Henden at
     ftp:/$\!$/ftp.nofs.navy.mil/pub/outgoing/aah/sequence/j2009.dat.
   \label{calib}
   \end{table}

\end{document}